\documentclass[a4paper,11pt]{article}
\usepackage{jinstpub} 
\usepackage{lineno}

\title{Coupler RF kick and emittance optimization  of the SHINE injector\footnote{This work was supported by the CAS Project for Young Scientists in Basic Research (YSBR-042), the National Natural Science Foundation of China (12125508, 11935020),Program of Shanghai Academic/Technology Research Leader (21XD1404100), Shanghai Pilot Program for Basic Research – Chinese Academy of Sciences, Shanghai Branch (JCYJ-SHFY-2021-010), the Youth Innovation Promotion Association CAS, China (No. 2021282), and the National Natural Science Foundation of China (No. 12205356).}}

\author[a]{Junjie Guo,}

\author[b,2]{Duan Gu,}
\author[b]{Zenggong Jiang,}
\author[b]{Zhen Wang,}
\author[b]{Meng Zhang,}
\author[b]{Qiang Gu,}
\author[b,2]{and Haixiao Deng\note{Corresponding author.}}

\affiliation[a]{Zhangjiang Laboratory, Shanghai 201210,  China}
\affiliation[b]{Shanghai Advanced Research Institute, Chinese Academy of Sciences, Shanghai 201210, China}
\emailAdd{gud@sari.ac.cn,denghx@sari.ac.cn}

\abstract{Coupler RF kick due to the asymmetric structure caused by the coupler, is more likely to lead to emittance growth in the SHINE injector with low beam energy. The calculation of coupler RF kick and resulting emittance dilution has been studied in detail in the literature. 
In this paper, a novel approach is provided that a lossy material is placed on the surface of the superconducting cavity to approximate the $Q_0$ of the TESLA cavity, and a frequency solver of CST is used to simulate the electromagnetic field distribution, which is used to calculate coupler RF kick, and calibrated against the results of CST Particle Tracking Studio with a good agreement. In order to minimize the emittance growth of SHINE injector, a  1.3 GHz symmetric twin-coupler cavity is adoped in the single-cavity cryomodule, and the rotational angle and permutation of the 8 cavities in the 8-cavities cryomodule is optimized. Ultimately, the optimized emittance is lower than the design parameter.}


\begin{document}
\maketitle
\flushbottom

\section{Introduction}

In recent years, high-repetition-rate XFEL based on superconducting Linac draws increasing attention due to its
unique capability of providing Ångström performance at high average power~\cite{b1}. Several facilities have been built in the
world such as Eu-XFEL~\cite{b2,b3} and LCLS-II~\cite{b4}. The Shanghai High Repetition rate XFEL and Extreme light facility (SHINE)~\cite{b5,b6,b7} is an x-ray FEL facility based on an 8 GeV continuous-wave superconducting RF linac, 3 undulator lines, 3 X-ray beamlines and 10 experimental stations, which has started its construction from april of 2018. It currently under construction at Zhangjiang Hi-Tech Park, close to the Shanghai Synchrotron Radiation Facility(SSRF)~\cite{b8} and the Shanghai soft X-ray free-electron laser(SXFEL)~\cite{b9}, aims to deliver FEL pulses between 0.4 keV and 25 keV with repetition rate up to 1 MHz, which enable research in various fields, including biology, chemistry, and material science. To meet the requirements of FEL performance, the linac of SHINE should provide 8 GeV high brightness electron beam up to 1 MHz, with normalized emittance of $\textless$ 0.5 µm at nominal 100 pC and peak current over 1500 A.  In particular, a high brightness injector is one of the key components of the SHINE facility. SHINE injector consists of a NC photocathode gun generating the electron bunch with 100 pC charge, a buncher, three solenoids, two accelerator cryomodule to boost the beam over 100 MeV with 12 A peak current~\cite{b10}.

The acceleration unit in SHINE injector is a 1.3 GHz 9-cell TESLA superconducting cavity~\cite{b11,b12} installed in the cryomodule, which is usually equipped with two higher-order-mode couplers (HOMC)~\cite{b13} to couple out the higher-order-mode~\cite{b14} in the cavity and a fundamental-power coupler (FPC)~\cite{b15} to feed the cavity with the RF acceleration field. However, these coupler structures break the axisymmetric structure of the cavity,  giving rise to relevant transverse multipole field components of the axial electromagnetic field in the coupler region. Therefore, when the electron bunch passes through the cavity along the axis of the cavity, it will still be subjected to the transverse kick force brought by these transverse electromagnetic field components, and we call the transverse kick force due to the coupler structure: coupler RF kick. In the SHINE injector, due to the low beam energy, the emittance growth caused by coupler RF kick is more serious, so it is necessary to evaluate the multipole field component and the coupler RF kick of the TESLA cavity in SHINE injector, and optimize the layout and structure of the superconducting cavity in the cryomodule to reduce the emittance growth. 

One of the difficulties is accurately to simulate the electromagnetic field distribution and to calculate the coupler RF kick.  Previously, numerous published studies~\cite{b16,b17,b18,b19,b20} describe using the eigenmode solver of CST~\cite{b21} or HFSS~\cite{b22} to simulate the traveling-standing wave field in a cavity and to calculate coupler RF kick. However, this method is complex and fails to take the loss on the surface and the $Q_0$ of the superconducting cavity into account. The present study aims to accurately solve for the electromagnetic field distribution and coupler RF kick using a novel approach in this paper. In this novel approach, a lossy material is placed on the surface of the superconducting cavity to approximate the $Q_0$ of the TESLA cavity, and a frequency solver of CST is used to simulate the electromagnetic field distribution. Moreover, the calculated result of coupler RF kick were calibrated against the result of CST Particle Tracking Studio, and confirmed to be in good agreement.

In the end, considering the coupler RF kick caused by the asymmetric coupler, a symmetric twin-coupler cavity will be adopted in the single-cavity cryomodule to eliminate the dipole mode component, and the rotational angle $\&$ permutation of the 8 cavities can be optimized to counteract the coupler RF kick in the 8-cavities cryomodule, which ultimately result in the lower emittance at the end of SHINE injector.

This paper is organized in the following way: it begins with an introduction to the layout of the SHINE injector and the structure of the main acceleration unit 1.3 GHz 9-cell TESLA superconducting cavity in Section.$\mathbf{\ref{sec:II}}$. Section.$\mathbf{\ref{sec.III}}$ provides the multipole field expansion coefficient $E_{m}/E_{0}$ at the coupler. Electromagnetic field distribution on axis are obtained by the CST frequency domain solver, and then coupler RF kick is calculated in Section.$\mathbf{\ref{sec.IV}}$. Section.$\mathbf{\ref{sec.V}}$ describes the scheme to optimize the  emittance growth of SHINE injector, where a symmetric twin-coupler cavity is adopted in the single-cavity cryomodule and the rotational angle and permutation of 8 TESLA cavities are optimized in the 8-cavities cryomodule. Finally, the full text is summarized in the section.$\mathbf{\ref{sec.VI}}$.
\begin{figure*}[!htb]
\includegraphics[width=\hsize]{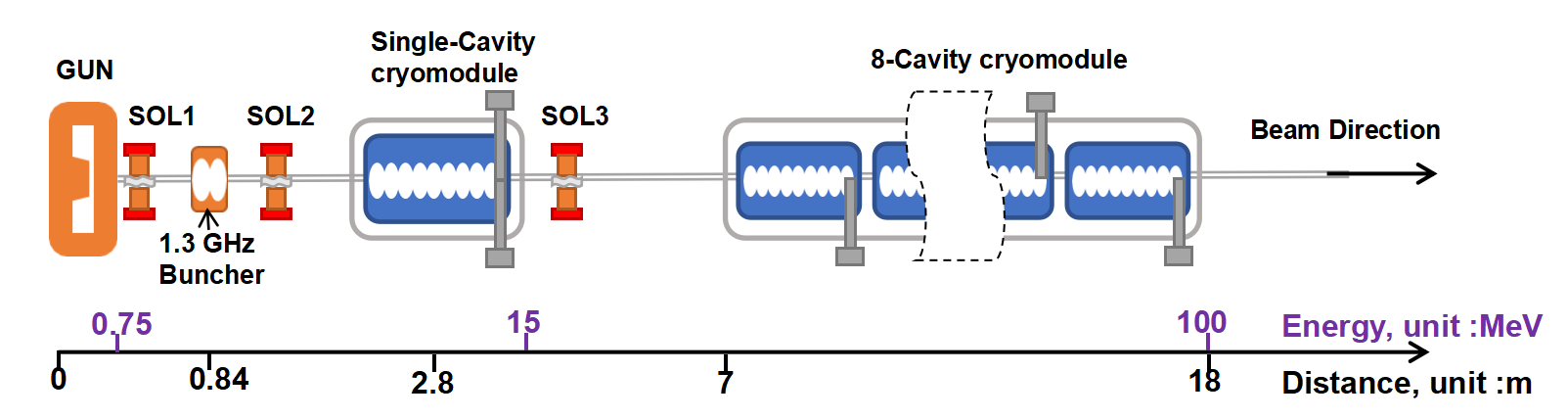}
\caption{The layout of SHINE injector.(GUN: 162.5 MHz VHF gun; SOL 1$\&$2$\&$3:  three movable solenoids; 1.3 GHz Buncher: a normal conducting RF buncher; single-cavity cryomodule: injCM01; 8-cavities cryomodule: injCM02.)}
\label{fig1}
\end{figure*}

\section{The layout of SHINE injector and geometry of TESLA cavity}\label{sec:II}

The simplified schematic layout of the SHINE injector, consisting of a VHF gun~\cite{b23,b24} operating at  162.5 MHz, three solenoids, a 1.3-GHz 2-cell normal-conducting RF buncher for electron bunch compression,  a 1.3-GHz SRF single-cavity cryomodule (injCM01) and a TESLA 1.3-GHz SRF 8-cavities cryomodule (injCM02), is shown in Fig.\ref{fig1}. The electron bunch are generated by the driving laser in the VHF Gun,
which energy is 0.75 MeV at the exit of the gun. Aiming to achieve lower emittance, the transverse space charge effect of the electron bunch need to be compensated by three movable solenoids(SOL 1,2,3) located along the beamline. The 1.3 GHz normal conducting RF buncher is used to compress the electron bunch about 3 times from about 30 ps to about 10 ps (FWHM), with the peak current of about 10 A after the buncher. The electron bunch were accelerated to 15 MeV by the first injCM01, and 100 MeV by the injCM02. The laser heater system downstream of the injCM02 generates an uncorrelated energy spread in the electron bunch to suppress the micro-bunching instabilities. Finally, the electron bunch would be focused by quadrupoles and transmitted to the downstream main accelerator. Table.\ref{tab1} summarizes the designed electron bunch parameters of the SHINE injector.

\begin{figure}[!ht]
    \centering
    \includegraphics[width=0.7\linewidth]{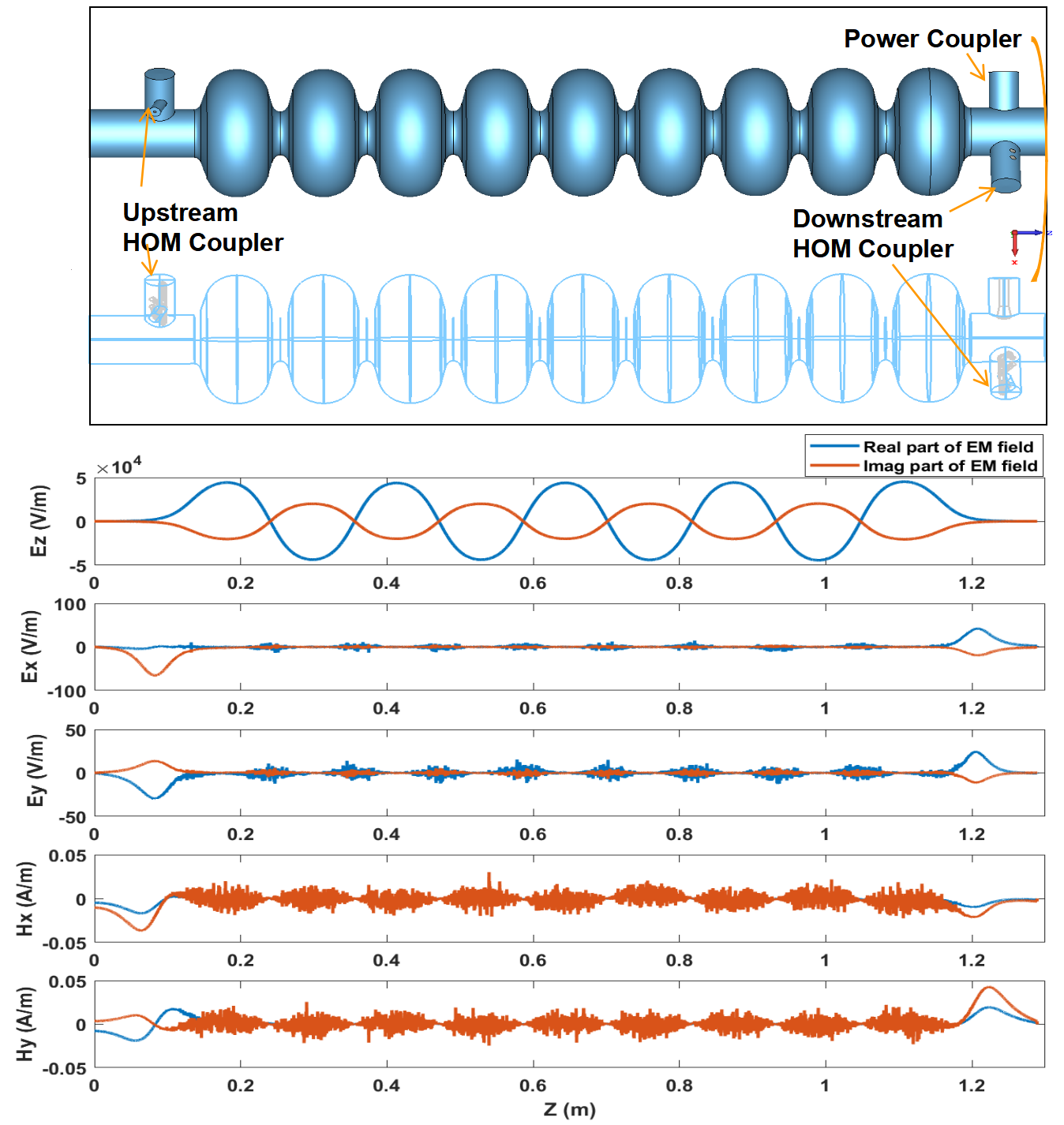}
    \caption{Longitudinal cross section(top plot) and electromagnetic field(bottom plot) on the axis of a 1.3 GHz 9-cell TESLA superconducting cavity.}
    \label{fig2}
\end{figure}

The acceleration unit in the injector cryomodule is a TESLA-type 9-cell standing wave structure \cite{b16} of about 1 m length whose fundamental accelerating mode resonates at 1.3 GHz, whose cavity structure and electromagnetic field distribution on the axis are shown in Fig.\ref{fig2}.  The significant advantage of superconducting niobium cavities as compared to normal conducting cavities is their low surface resistance of about a few $n\Omega$ when cooled by liquid helium to a few K, which results in high wall currents and low heat losses.
In order to define a measure of this quality of the cavity, the so-called quality factor, which includes the intrinsic quality factor $Q_0$, the external quality factor $Q_{ext}$ and the loaded quality factor $Q_L$, the relationship between these three is defined as
\begin{equation}
\frac{1}{Q_L}=\frac{1}{Q_0}+\frac{1}{Q_{\mathrm{ext}}} = \frac{P_ {diss}}{\omega_0 W}+ \frac{P_{e x t}}{\omega_0 W}
\label{equ1}
\end{equation}

where $W$ is the stored energy, $\omega_0$ is the resonance frequency, $P_{diss}$ the dissipated power of the cavity and $P_{ext}$ the dissipated power in external devices, for superconducting cavities $Q_{ext} \ll Q_{0}$. The intrinsic quality factors of normal conducting cavities are in the range of $10^4$-$10^5$, while for TESLA cavities $Q_{0}>10^{10}$. Compared to the room temperature cavity with low  $Q_0$, the heat losses of the high $Q_0$ superconducting cavity is very small, so that the operation of electron bunch  with high repetition frequency can be realized.  $Q_0$ means that the ratio of the stored energy in the cavity to the powe losses in the cavity wall, while $Q_{ext}$ implies that energy is not only dissipated in the cavity walls but also extracted through the fundamental power coupler and dissipated in an external load. The fundamental power coupler downstream of the TESLA superconducting cavity transmits power from an external power source through the coaxial conductor to the inside of the cavity, where factors such as the structure and position of the coupler, and the insertion depth of the inner conductor determine the $Q_{ext}$. Further, these factors also affect the asymmetry and transverse field distribution of the cavity.

To meet the needs of the SHINE project, the fundamental power coupler was modified and optimized based on the TTF-III FPC design, in which the antenna of the cold part was shortened by 8.5 mm in order to obtain a higher external quality factor $Q_{ext}$ with a value of $4.12 \times 10^7$\cite{b15}, under the following working conditions: the average linac electron bunch current is 0.3 mA, the nominal accelerating gradient $E_{acc}$ of the 1.3 GHz superconducting cavity is 16.0 MV/m, the cavity intrinsic quality factor $Q_0$ is $2.7 \times 10^{10}$, and the frequency detuning $\delta f$ caused by the peak microphonics effect is 10 Hz~\cite{b26}.

Due to the presence of the asymmetric coupler, the axial symmetry of the cavity is broken, and the transverse component of the electromagnetic field at the coupler is excited,  as shown in Fig.\ref{fig2} (bottom plot). This electromagnetic field distribution is simulated by the CST frequency solver and will be described in detail in Section.$\mathbf{\ref{sec.IV}}$.
Nonaxisymmetric structure excites multipole field components, such as the dipole, quadrupole, and octopole, which kick the electron bunch time dependently, leading to the emittance growth.

    \begin{table}[!htb]
		\caption{The Design Parameters of SHINE Injector}
		\label{tab1}
		\centering
		\begin{tabular}{lccc}
			\hline
			Parameter & Value \\
			\hline
			Energy & 100 MeV  \\
			Charge/bunch & 100 pC \\
			Bunch length (RMS) & 1 mm \\
			Peak current & 12 A\\
                Normalized slice  emittance (RMS, 95$\%$) & 0.4 $\mu$m \\
			\hline
		\end{tabular}
    \end{table}

\section{Multipole field expansion coefficient}\label{sec.III}

The geometric influence of the coupler on the RF structure leads to the asymmetry of the structure, thereby introducing the asymmetry of the electromagnetic field, which will excite the multipole field components that cause the emittance growth, such as Dipole mode, Quadrupole mode, Sextupole mode, and Octopole mode~\cite{b27}. Such electromagnetic field can be expanded by fourier series according to multipole field components. The TM-mode longitudinal electric field in the superconducting cavity can be approximated according to radial and azimuth changes and expanded into the following formula:
\begin{figure}[!htb]
   \centering
   \includegraphics*[width=0.7\linewidth]{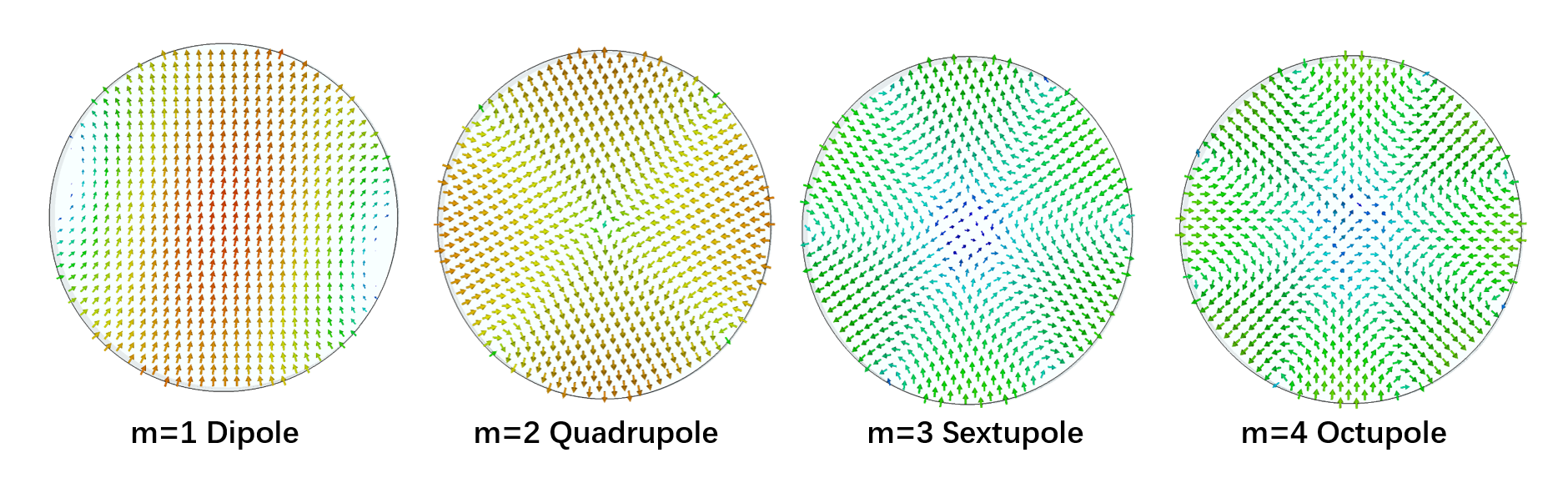}
   \caption{ Schematic drawing of each multipole mode. }
   \label{fig3}
\end{figure}

\begin{equation}
    E_z(r, \theta, z) \sim \sum_m E_{z m}(z) r^m \cos (m \theta)
    \label{equ2}
\end{equation}

As shown in Equ.(\ref{equ2}), the 0th component m=0 of the multipole field is the monopole mode, the 1st
component m=1 is the dipole mode, the 2nd component m=2 is the quadrupole mode, the 3rd component m=3 is
the sextupole mode, and the 4th component m=4 is octopole mode, and so on. Equ.(\ref{equ2}) can be expanded to order 4 to obtain the following Equ.(\ref{equ3}):
\begin{equation}
\begin{aligned}\label{equ3}
E_z  \approx \sum_m E_{z m}(z) r^m \cos (m \theta) 
=E_{z 0}(z) r^0 \cos (0 * \theta)+E_{z 1}(z) r^1 \cos (\theta)\\ 
+E_{z 2}(z) r^2 \cos (2 * \theta)+E_{z 3}(z) r^3 \cos (3 * \theta)+E_{z 4}(z) r^4 \cos (4 * \theta) \ldots
\end{aligned}
\end{equation}

which is expanded over radial $r$ and azimuth $\theta$. Fig.\ref{fig3} shows transverse electric field distribution of each multipole mode schematically, which gives us a visualization of each mode.

\begin{figure}[!htb]
   \centering
   \includegraphics*[width=0.7\linewidth]{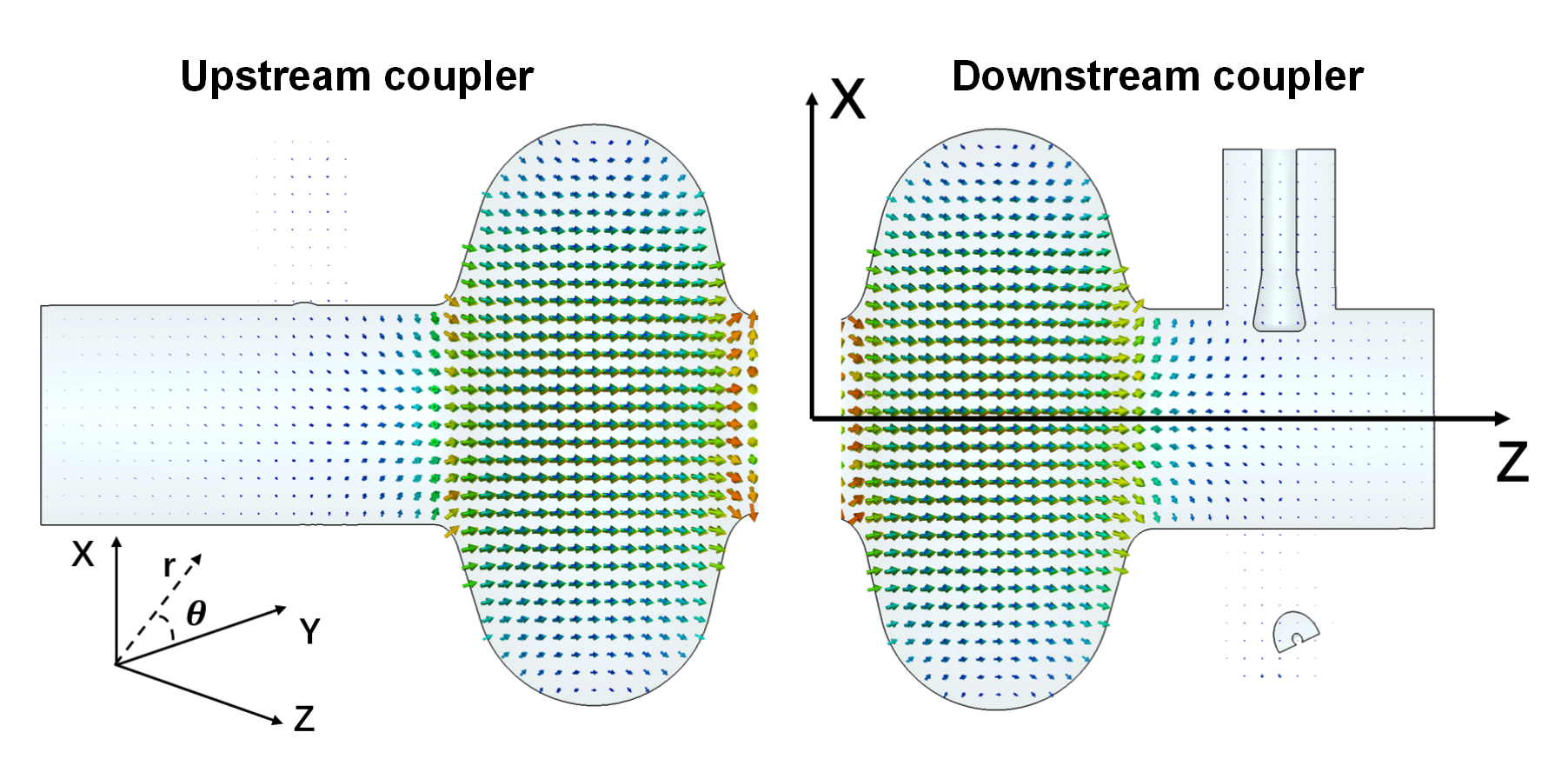}
   \caption{ Electromagnetic field distribution in the upstream and downstream coupler.  }
   \label{fig4}
\end{figure}

\begin{figure}[!htb]
   \centering
   \includegraphics*[width=0.7\linewidth]{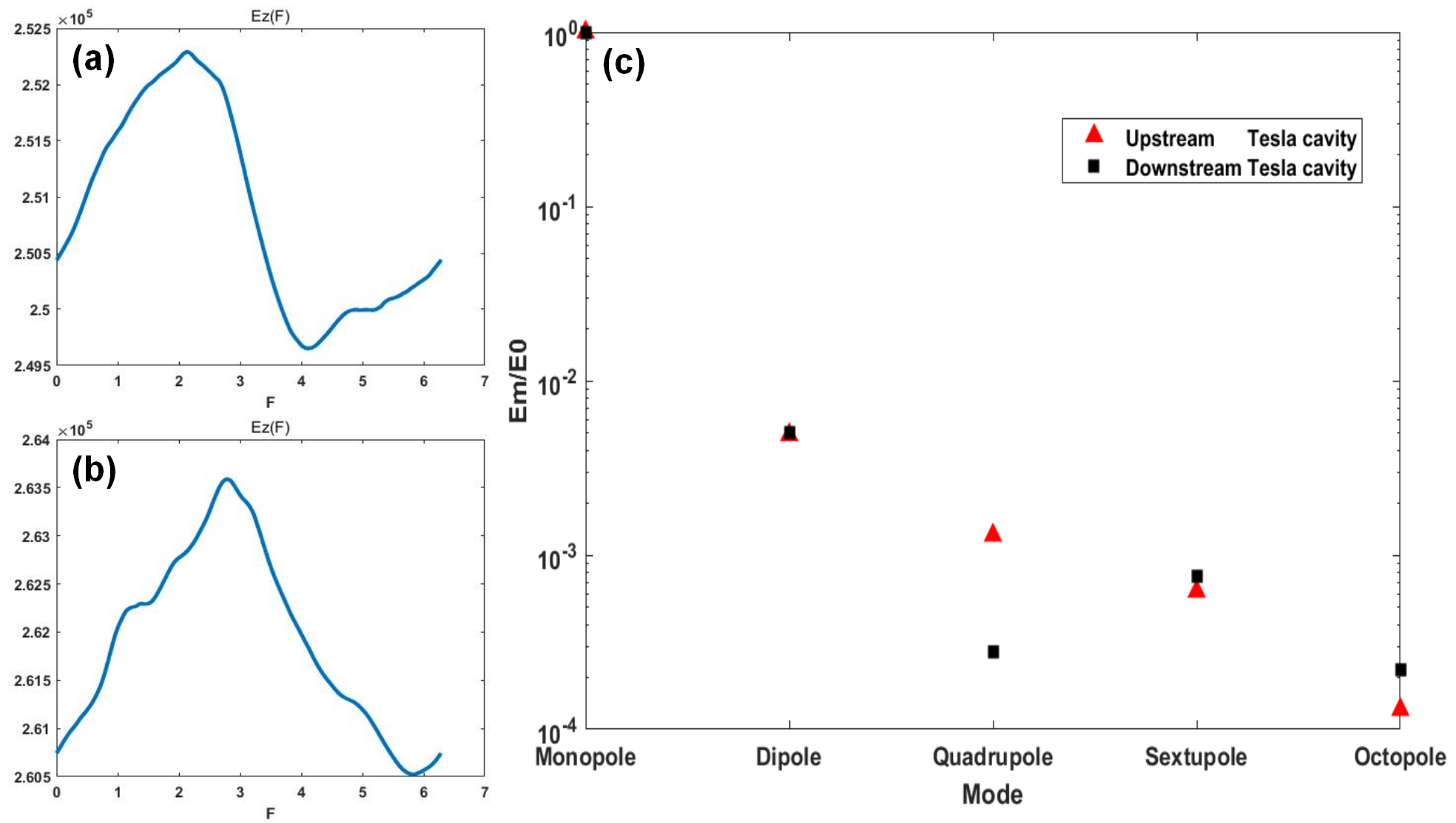}
   \caption{Ez field at r = 3 mm upstream coupler (a) and downstream coupler (b) of TESLA cavity, normalized multipole mode expansion coefficient $E_{m}/E_{0}$ (c).}
   \label{fig5}
\end{figure}

The eigenmode solver of the CST was used to simulate the 3D electromagnetic field distribution at the upstream and downstream couplers of the TESLA cavity, as shown in Fig.\ref{fig4}. The central axis of the cavity is the Z axis, the
direction of the fundamental-power coupler is the X-axis, and the azimuth angle $r-\theta$ in the X-Y plane is defined.
The field distribution shown in Fig.\ref{fig4} is a plot of the electromagnetic field in the TESLA cavity near the coupler when the acceleration phase is 0 degrees, which includes multipole mode information such as monpole mode, dipole mode, quadrupole mode, sextupole mode, and octopole mode.
We need to decompose these multipole modes into each individual mode to obtain their amplitude
and phase information. Here we use the fast fourier transform (FFT) algorithm to obtain the magnitude and phase information of these multipole modes. In this case, the FFT is performed with respect to spatial coordinates (azimuthal angle in the cavity) rather not carried out in time coordinates.
In order to perform a fast fourier transform for the higher order modes, we take the Ez distribution along the $\theta$ direction at r = 3 mm, middle of the upstream and downstream coupler in Fig.\ref{fig4}, as shown in Fig.\ref{fig5} (a) and (b).

Afterwards, the $E_z$ curves of Fig.\ref{fig5} (a) and (b) are processed to the fourier coefficients $E_{m}$$(a_{m}r_{0}^{m})$ for each of the multipole modes by fast fourier transform, as shown in Fig.\ref{fig5} (c) and Table.\ref{tab2}.
From Fig.\ref{fig5} (c) and Table.\ref{tab2}, it can be seen that the electromagnetic field in the upstream and downstream couplers of the TESLA cavity is dominated by the dipole mode.  As can be seen from Fig.\ref{fig5} (a) and (b), the RF field is mainly in the shape of a cosine function of a period, that is, it is mainly the influence of the dipole mode field.
Therefore, it can be associated that the dipole mode is the dominant multipole mode for structures with only one coupler, the quadrupole mode is dominant for structures with two symmetric couplers, and the octopole mode is the dominant one for structures with four symmetric couplers.
Although there are two couplers in the downstream coupler of the TESLA cavity, they are not symmetrical, and the combined effect is one coupler. In this way, there is a strong dipole mode at the upstream and downstream couplers of the TESLA cavity, which will produce a strong transverse kick force on the electron bunch. 

\newcommand{\tabincell}[2]{\begin{tabular}{@{}#1@{}}#2\end{tabular}}  

	\begin{table}[!htbp]
		\caption{Normalized multipole mode expansion coefficient $E_{m}/E_{0}$
at TESLA cavity upstream and downstream coupler $Q_e$ = $4.12 \times 10^{7}$, $r_0$ = 3 mm. }
		\label{tab2}
		\centering
		\begin{tabular} {ccc}
			\hline
			Mode & \tabincell{c}{Upstream \\ coupler} & \tabincell{c}{Downstream \\ coupler}  \\ 
			\hline
			$E_{0}/E_{0}$ ($a_{0}/a_{0}$),monopole & 1 & 1   \\
			$E_{1}/E_{0}$ ($a_{1}r_{0}/a_{0}$),dipole & {\boldmath$4.9 \times$ $10^{-3}$} & {\boldmath $5.0 \times 10^{-3}$}  \\
			$E_{2}/E_{0}$ ($a_{2}r_{0}^{2}/a_{0}$),quadrupole & $1.3 \times 10^{-3}$ &$2.8 \times 10^{-4} $\\
			$E_{3}/E_{0}$ ($a_{3}r_{0}^{3}/a_{0}$),sextupole & $6.2 \times 10^{-4}$& $7.6 \times 10^{-4}$\\
			$E_{4}/E_{0}$ ($a_{4}r_{0}^{4}/a_{0}$),octopole &$1.3 \times 10^{-4}$& $2.2 \times 10^{-4}$\\
			\hline
		\end{tabular}
	\end{table}

\section{Calculation of EM-field and Coupler RF Kick}\label{sec.IV}
While the previous subsection was only the multipole field expansion of the field distribution at the middle position of the coupler, we would like to know the effect of the electromagnetic field in the whole coupler region and the transverse kick that the electron bunch receives when it passes through the 1.3 GHz fundamental mode at a certain synchronization phase.

\begin{figure}[!htb]
   \centering
   \includegraphics*[width=0.7\linewidth]{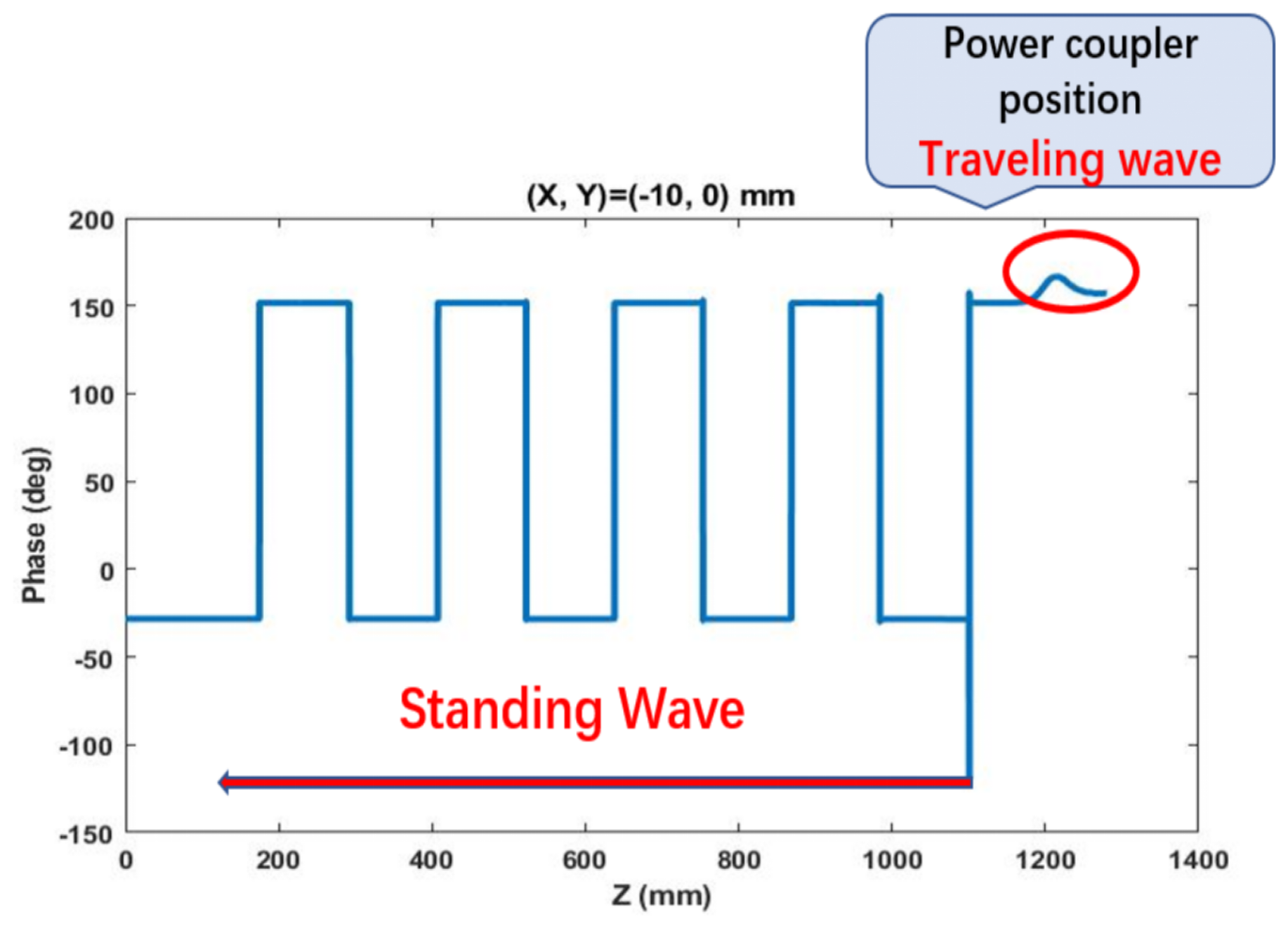}
   \caption{Phase diagram of traveling and standing wave field in a TESLA cavity.}
   \label{fig6}
\end{figure}

The TESLA cavity is a 9-cell standing-wave accelerated structure operating in $\mathbf{TM_{010;\pi}}$ mode, where the standing-wave field is formed by the superposition of the forward traveling-wave field from the downstream FPC and the backward traveling-wave field reflected from the upstream. 
The standing-wave field obtained in this way can continuously accelerate electrons and improve acceleration efficiency. The standing-wave field is only inside the 9-cell cavity, but the electromagnetic field at the downstream FPC is still a traveling-wave field, as shown in the phase diagram of Fig.\ref{fig6}.  
For calculating the coupler RF kick, the most important step is to obtain the traveling-standing wave field distribution. 

\begin{figure}[!htb]
   \centering
   \includegraphics*[width=0.7\linewidth]{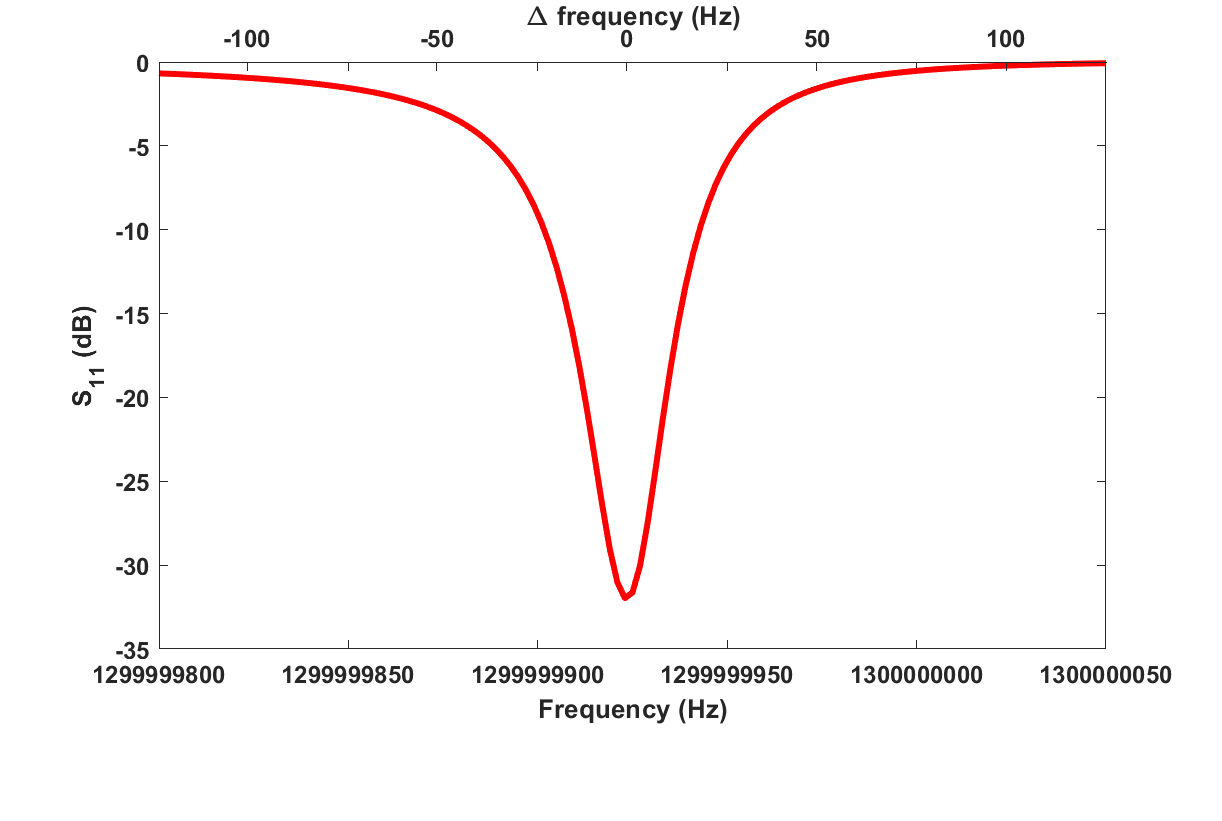}
   \caption{$S_{11}$ parameter curves versus
frequency for FPC of TESLA cavity simulated by the CST frequency solver.}
   \label{fig7}
\end{figure}

In previous studies, to reconstitute this incoming traveling wave, we can use results from a computer code like eigenmode solver of CST. The calculations have to be performed for two different boundary conditions at the end of the fundamental-power coupler: perfect electric wall and perfect magnetic wall. Then one can reconstitute the incoming traveling wave by combining the two standing wave solutions.
However, this method does not take into account the surface loss and the value $Q_0$ of superconducting cavities. In order to simulate the traveling-standing wave field of the coupler region and inside of cavity, we use the CST frequency solver with the input port on the fundamental power coupler, and set the loss material on the outer surface of the cavity to simulate the cavity wall loss and $Q_0$. In order to enable most of the input power from the fundamental power coupler into the cavity with very little reflected power, the parameters and area of the lossy material are constantly optimized, and here we have a small piece of copper lossy metallic material at the middle cell of the cavity. Special attention is paid to the fact that since the $Q_{0}$ of the superconducting cavity is very high, and the bandwidth of its frequency is very narrow, the frequency range of the $S_{11}$-parameters simulated by the CST frequency solver has to be set very small so that the resonant frequency $f_0$ $\approx$ 1.3 GHz can be found, as show in Fig.\ref{fig7}. 

\begin{figure*}[!hbt]
\centering
\includegraphics[width=\linewidth]{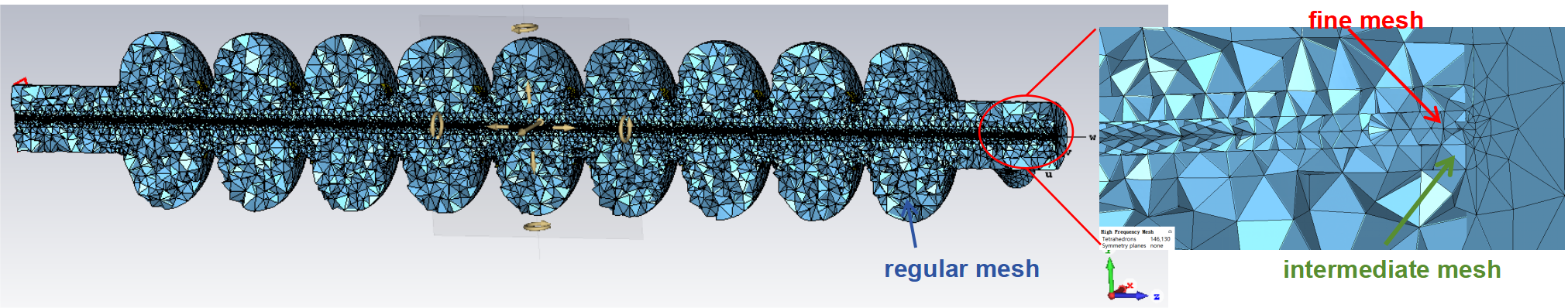}
\caption{The three-zone mesh for CST used in order to improve the field approximation near the axis.}
\label{fig8}
\end{figure*}

In order to accurately estimate the electromagnetic field, a mesh densification method \cite{b28} is utilized by the CST simulation with a non-uniform tetrahedral mesh, which include a unique three-zone mesh (regular mesh, intermediate mesh, and fine mesh, as shown in Fig.\ref{fig8}) at the axis of the cavity to enhance the accuracy of the field approximation near the axis. Here, a vacuum cylinder with r = 2 mm is used as the intermediate mesh and fine mesh, and a vacuum cylinder with r = 5 mm is set up as the intermediate mesh to match the fine mesh near the axis and the regular mesh of the rest of the cavity. 
Finally, the simulated electromagnetic fields $\left(\mathbf{E}_z, \mathbf{E}_x, \mathbf{E}_y, \mathbf{H}_x, \mathbf{H}_y\right), $ along the cavity axis are plotted in Fig.\ref{fig2} (bottom plot) for real part and imaginary part. From Fig.\ref{fig2}, it is observed that the electromagnetic field in the middle of the cavity has a spurious transverse component, probably due to any misalignment of the mesh elements with respect to the axis resulting in a non-zero transverse projection of the longitudinal component, but none of this affects the result of integrating the field.

When a charged particle with a velocity $v_{z}$  and a charge of $q$ passes through the RF cavity along a trajectory z, the transverse Lorentz force is felt by the electromagnetic field in the cavity:
\begin{equation}
F_{\perp}= q[\overrightarrow{\mathbf{E}}_{\perp}+(v_{z} \times \mu_{0}\overrightarrow{\mathbf{H}}_{\perp})] 
    \label{equ4}
\end{equation}

where $\overrightarrow{\mathbf{E}}$ and  $\overrightarrow{\mathbf{H}}$ are  vector electromagnetic field, $\mu_{0}$ is the magnetic permeability. $\overrightarrow{\mathbf{E}}$ can be decomposed further: 

\begin{equation}
\begin{split}
\overrightarrow{\mathbf{E}}=E_{0} e^{i\omega t+\varphi_{0}} \cdot e^{ikz}    \\
=(E_{Re}+i \cdot E_{Im})* (cos(kz)+i \cdot sin(kz))
\label{equ5}
\end{split}
\end{equation}

where the real and imaginary electromagnetic fields $E_{Re}$, $E_{Im}$ can be obtained by the CST frequency solver, as in Fig.\ref{fig2} (bottom plot).  
$E_{0}$ is the amplitude of the electromagnetic field corresponding to each coordinate,  $\varphi_{0}$ is the synchronous phase of the electron bunch and the peak accelerating voltage,  $\omega$ is the resonant frequency and $k$ is the wavenumber. In the same way, the vector magnetic field can be decomposed:
\begin{equation}
\begin{split}
\overrightarrow{\mathbf{H}}=H_{0} e^{i\omega t+\varphi_{0}} \cdot e^{ikz} \\
=(H_{Re}+i \cdot H_{Im})*(cos(kz)+i \cdot sin(kz))
\label{equ6}
\end{split}
 \end{equation}

where the real and imaginary electromagnetic fields $H_{Re}$ and $H_{Im}$ are obtained by the CST frequency solver, as in Fig.\ref{fig2} (bottom plot).  

After we have determined the time-harmonic electromagnetic field in the RF cavity and the synchronous phase $\varphi_{0}$ of the charged particles with the RF field, the effective voltage felt by the charged particles in the transverse (x, y) direction is 
\begin{equation}
\mathbf{V}_{\perp}(x, y)=\int d z\left[\overrightarrow{\mathbf{E}}_{\perp}(z)\pm v_{z} \times \overrightarrow{\mathbf{B}}_{\perp}(z)\right] e^{i \frac{\omega z}{c}}
\label{equ7}
\end{equation}

\begin{equation}
\overrightarrow{\mathbf{V}}_{x}=\int [\rvert \overrightarrow{\mathbf{E}}_{x}(z)-v_{z} \times \mu_{0} \overrightarrow{\mathbf{H}}_{y}(z)\rvert] \cdot e^{i (\frac{\omega  z}{c}+\phi_{0})} d z
\label{equ8}
\end{equation}

\begin{equation}
\overrightarrow{\mathbf{V}}_{y}=\int 
\left[\left(\overrightarrow{\mathbf{E}}_{y}(z)+v_{z} \times \mu_{0} \overrightarrow{\mathbf{H}}_{x}(z)\right] \cdot e^{i (\frac{\omega z}{c}+\phi_{0})} d z\right.
\label{equ9}
\end{equation}

where $c$ is the speed of light and  $\phi_{0}$ is the synchronous phase of the electron bunch with the radio frequency field. And the effective longitudinal accelerating voltage felt by charged particles as they pass through the TESLA cavity: 
\begin{equation}
   V_{\mathrm{z}}=\int E_{z}(z) \cdot e^{i (\frac{\omega z}{c}+\varphi_{0})} d z.
   \label{equ10}
\end{equation}

The normalized vector coupler RF kick is defined as the ratio of the effective transverse voltage to the effective longitudinal accelerating voltage:

\begin{equation}
{\mathbf{K}}_{x / y}={\mathbf{V}}_{x / y} / V_{\mathrm{acc}}
\label{equ11}
\end{equation}

\begin{figure}[!htb]
    \centering
    \includegraphics*[width=0.7\linewidth]{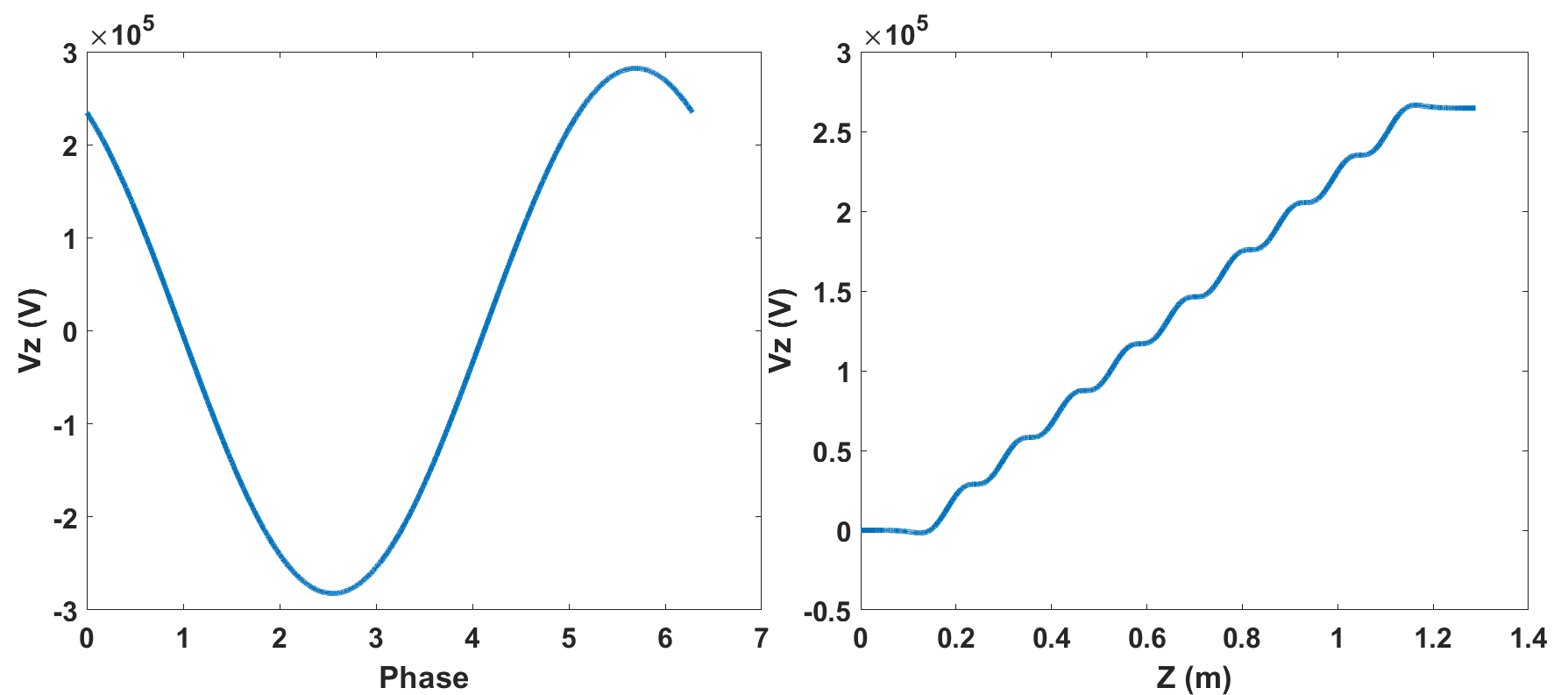}
    \caption{$V_{z}$ versus phase (Left) and $V_{z}$ at the peak acceleration phase along z (Right).}
    \label{fig9}
\end{figure}

\begin{figure*}[!htb]
\centering
\includegraphics[width=0.7\linewidth]{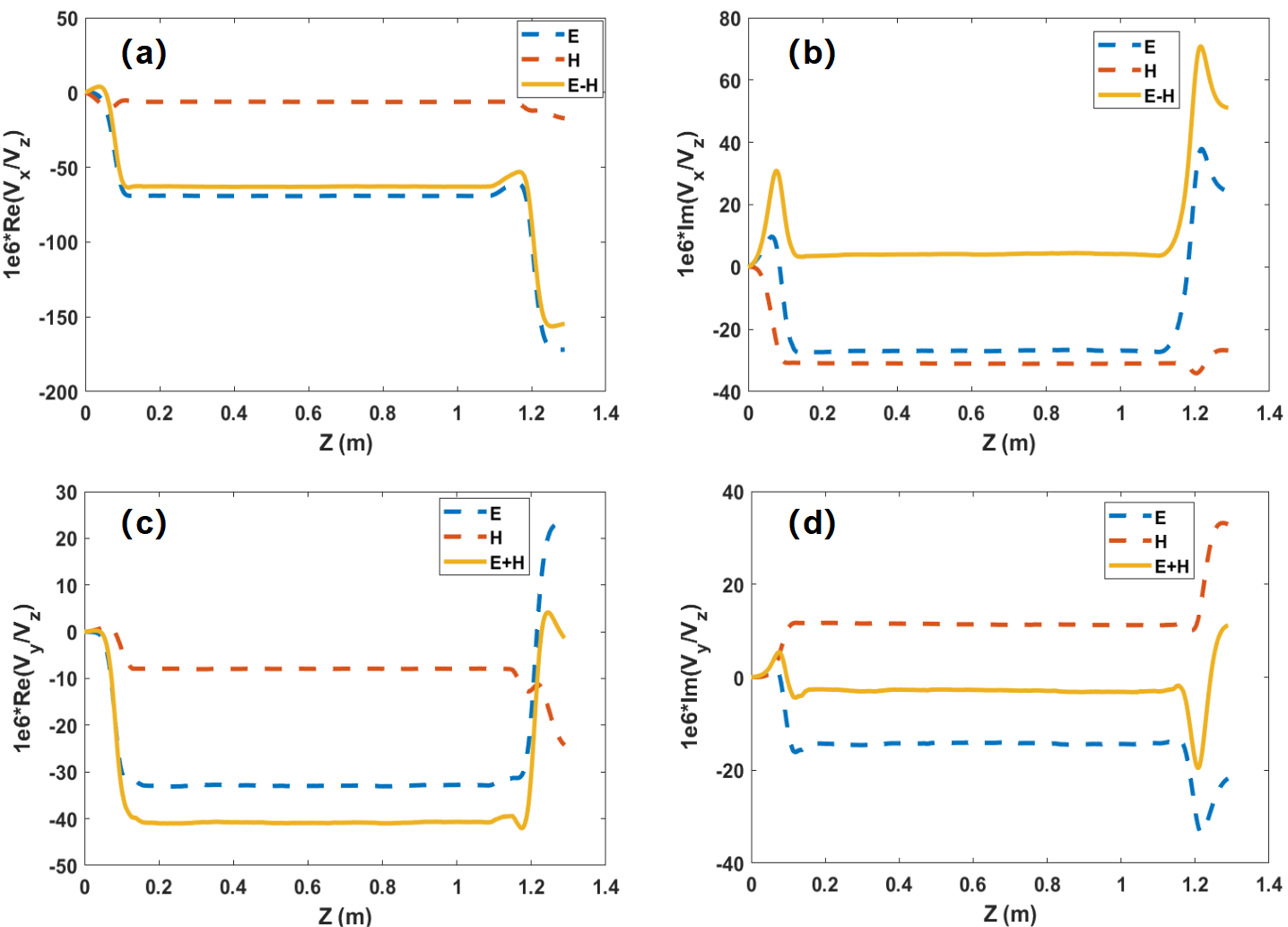}
\caption{ Real (a) and imaginary (b) parts of ${\mathbf{V}}_{x/z}$ and real (c) and imaginary (d) parts of ${\mathbf{V}}_{y/z}$ of the TESLA cavity}
\label{fig10}
\end{figure*}

According to the electromagnetic field distribution on the axis shown in Fig.\ref{fig2} and using Equ.(\ref{equ4}-\ref{equ10}), $V_{z}$ can be solved for the phase dependence, as shown in Fig.\ref{fig9} (left plot), which determines the peak phase of the longitudinal accelerating voltage $V_{z}$ to be 5.6087.
Taking its value as the synchronized phase of the electrons on the RF, i.e., the peak acceleration, this gives the $V_{z}$ along z, as in Fig.\ref{fig9} (right plot).
Similarly, the transverse $V_{x}$ and $V_{y}$ seen by the electron bunch can be derived, and substituting Equ.(\ref{equ11}) gives the coupler RF kick for the entire cavity region, as shown in Fig.\ref{fig10}. Detailed coupler RF kick of upstream coupler region, downstream coupler region and the entire Cavity region can be found in Table.$\ref{tab3}$.

    \begin{table}[!htb]
		\caption{Coupler RF Kick of upstream coupler region, downstream coupler region and the entire Cavity region.}
		\label{tab3}
		\centering
		\begin{tabular}{lccc}
			\hline
			$Q_e$=$4.12 \times 10^{7}$ & Upstream & Downstream &Full Kick \\
			\hline
			1e6*Kick X(Re) & -62.738 & -91.6577 &-154.39  \\
			1e6*Kick Y(Re) & -40.990 & 39.3553  & -1.6352 \\
			1e6*Kick X(Im) & 3.7640   &47.1559   &50.9199 \\
			1e6*Kick Y(Im) & 2.7264   &14.2892   &	17.0156\\
			\hline
		\end{tabular}
    \end{table}

\begin{figure}[!ht]
  \begin{minipage}[b]{1\linewidth}
    \centering
    \includegraphics[width=0.5\linewidth]{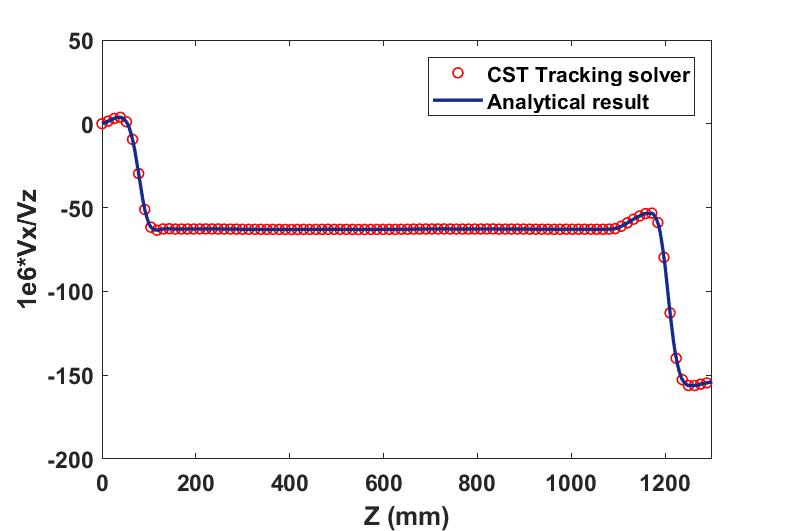}
    \label{fig:subfiga}
  \end{minipage}
  \hfill
  \begin{minipage}[b]{1\linewidth}
    \centering
    \includegraphics[width=0.5\linewidth]{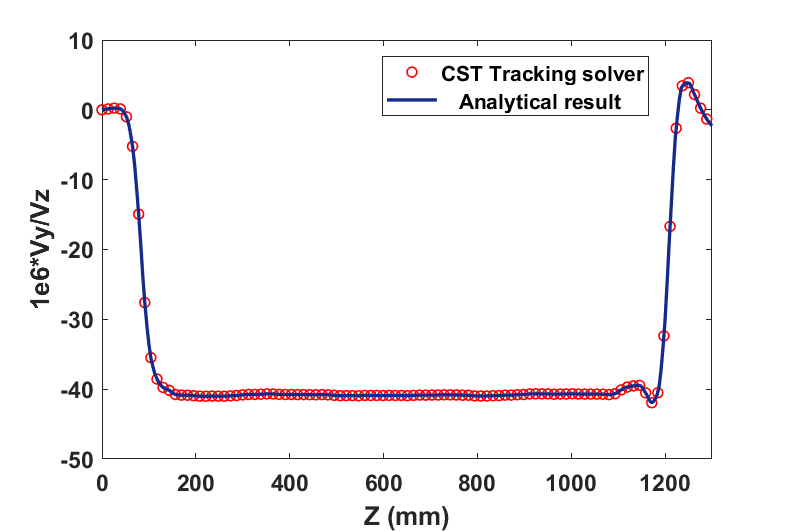}
    \label{fig:subfigb}
  \end{minipage}
  \caption{Comparison of ${\mathbf{V}}_{x / z}$ (top plot) and ${\mathbf{V}}_{y / z}$ (bottom plot) obtained by the CST Particle Tracking Studio with the results of the analytical formulas.}
  \label{fig11}
\end{figure}

The coupler RF kick calculated from analytical Equ.(\ref{equ4}-\ref{equ11}) can be verified using CST Particle Tracking Studio in the following steps: 
\begin{itemize}
    \item [(a)] loading the electromagnetic field solved by CST frequency solver into the external field of CST Particle Tracking Studio
    \item [(b)] scanning the phase to find the phase position with respect to the peak accelerating voltage $V_{z}$ at Fig.\ref{fig9}
    \item [(c)] setting up the parameters of a rigid point-charged particle source 
    \item [(d)] multiple beam position-momentum monitors are placed to obtain the real part of the ${\mathbf{V}}_{x / z}$ and ${\mathbf{V}}_{y / z}$
\end{itemize}
Finally, the results obtained using CST Particle Tracking Studio are in good agreement with those derived from analytical Equ.(\ref{equ4}-\ref{equ11}), as shown in Fig.\ref{fig11}.

\section{Emittance Optimization of the SHINE injector}\label{sec.V} 
\begin{figure}[!ht]
  \begin{minipage}[b]{1\linewidth}
    \centering
    \includegraphics[width=0.7\linewidth]{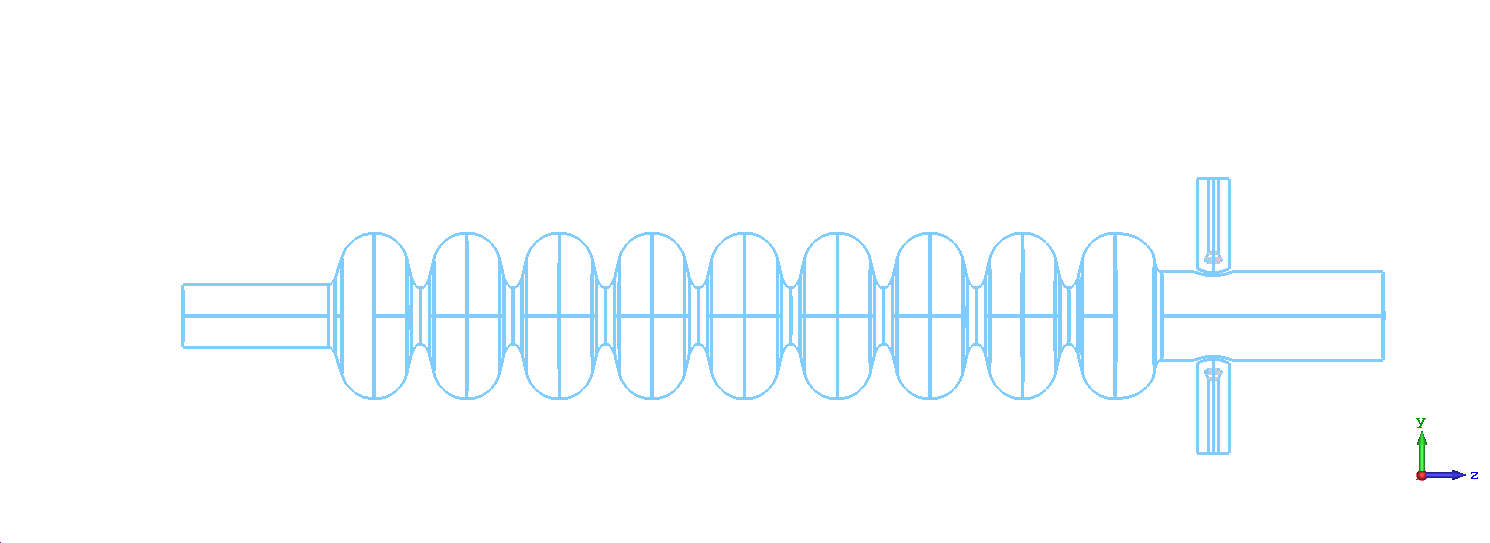}
    \label{fig:subfiga}
  \end{minipage}
  \hfill
  \begin{minipage}[b]{1\linewidth}
    \centering
    \includegraphics[width=0.7\linewidth]{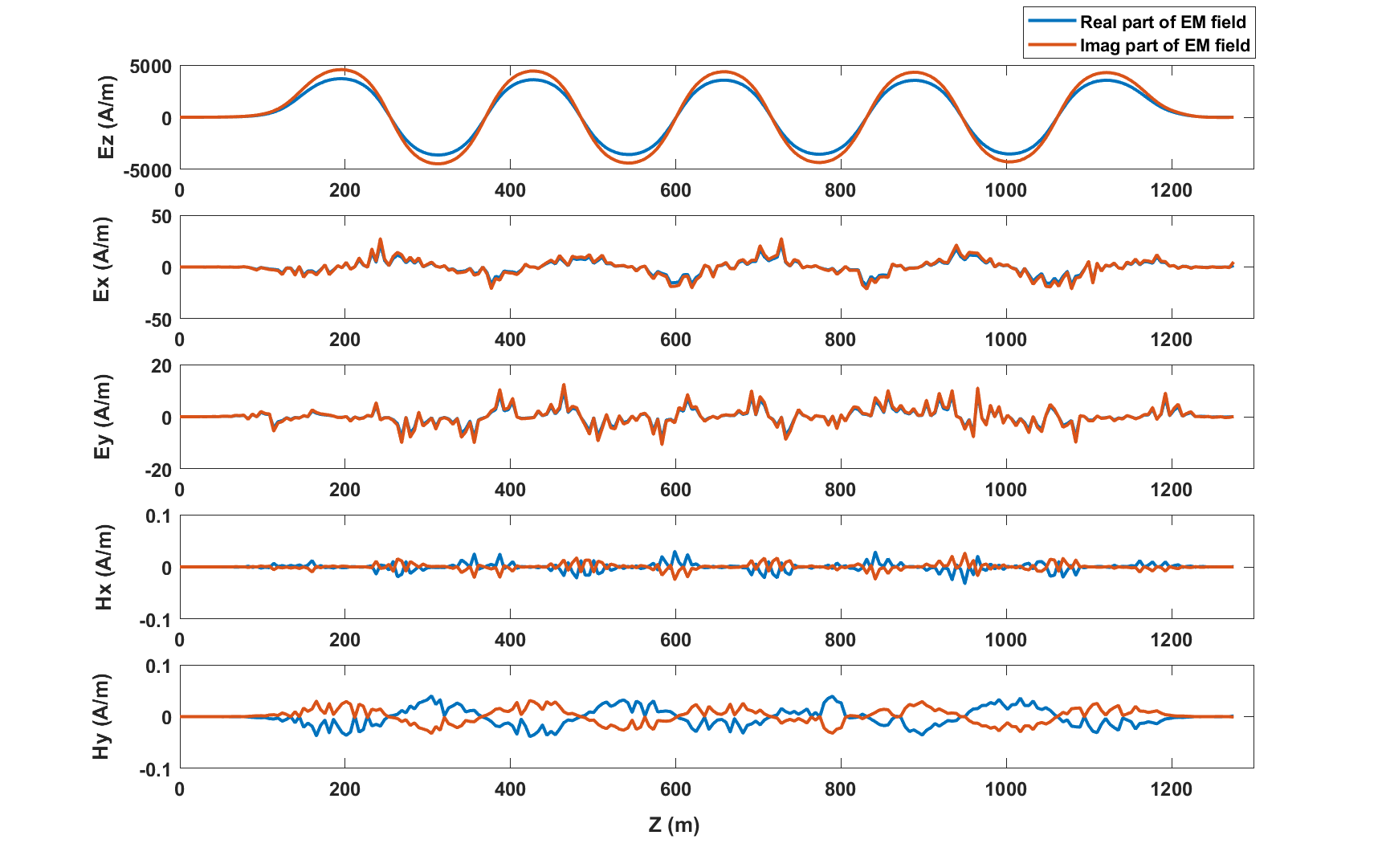}
    \label{fig:subfigb}
  \end{minipage}
  \caption{Structure (Top plot) and Electromagnetic fields (Bottom plot) on the axis of 1.3 GHz 9-cell symmetrical twin-coupler cavity.}
  \label{fig12}
\end{figure}
In the SHINE injector, the coupler RF kick results in a significant increase in electron bunch emittance due to the low electron bunch energy. In order to reduce the emittance growth of the SHINE injector, a symmetrical 1.3 GHz twin-coupler cavity can be used in the injCM01, and the rotation angle and permutation of the cavity in the injCM02 can be optimized to decrease the beam emittance.
Refer to the relevant contents in Sec.\ref{sec.III}, we infer that dipole mode of the structure with two symmetric couplers is no longer the dominant mode, such that the electromagnetic field distribution on the axis will not have a transversal dipole kick. Therefore, a symmetrical twin-coupler cavity is designed in the injCM01 of SHINE injector in the following steps: 
\begin{itemize}
    \item [1)] the upstream end of the symmetrical twin-coupler cavity only removes the higher-order mode coupler, keeping the other dimensions unchanged
    \item [2)]  this cavity still maintains 9 cells, with the dimensions of the middle 7 cells and the end cell of upstream remaining unchanged from the TESLA 9-cell cavity
    \item[3)] the downstream of the symmetrical twin-coupler cavity is changed from the TESLA 9-cell cavity in the following steps:
    \begin{itemize}
        \item [a)] two higher-order-mode couplers are removed
        \item[b)]  a fundamental-power coupler is installed opposite the original fundamental-power coupler to form a symmetrical structure
        \item[c)] the beam pipe radius is expanded to 55 mm and the dimensions of the end half-cell are changed accordingly, so that the higher-order mode with larger R/Q can be transmitted out.
    \end{itemize}
\end{itemize}

\begin{figure}[!htb]
   \centering
   \includegraphics*[width=0.7\linewidth]{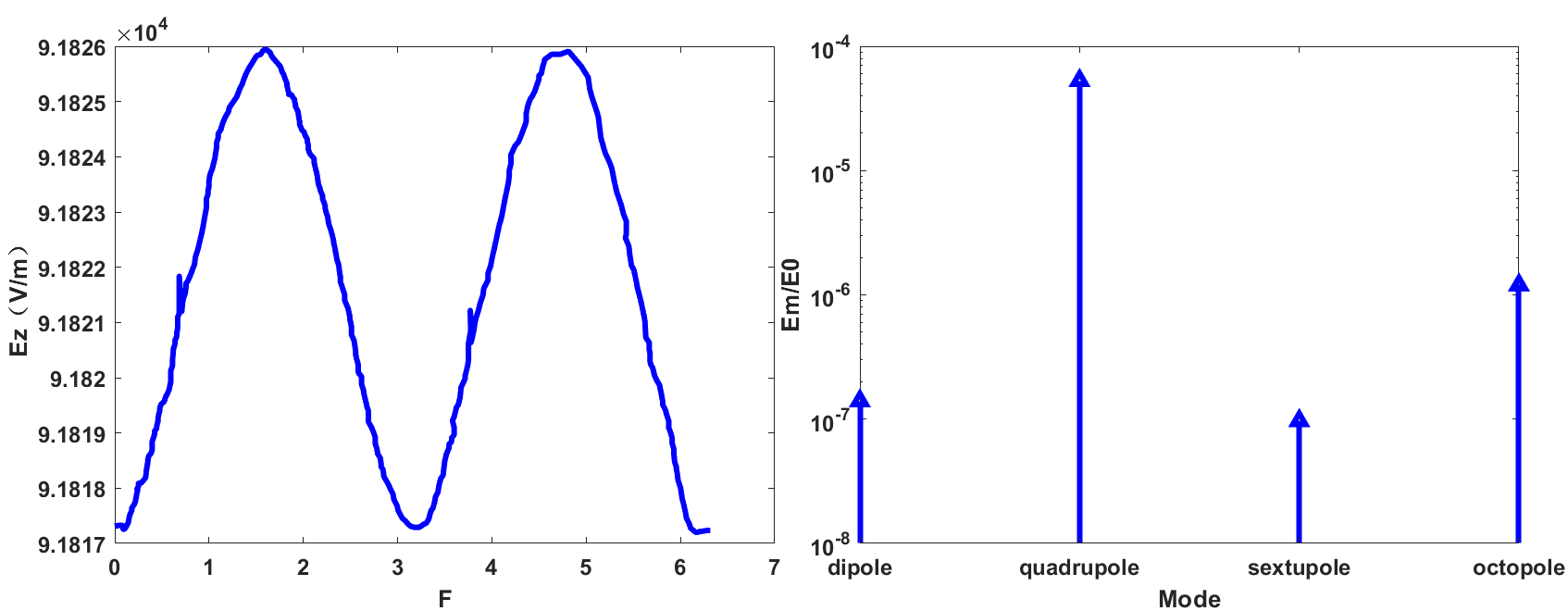}
   \caption{$E_z$ field at r= 3 mm downstream coupler (left plot) and normalized Fourier coefficient $E_{m}/E_{0}$ (right plot) of 1.3 GHz 9-cell  symmetrical twin-coupler cavity.}
   \label{fig13}
\end{figure}

Finally, geometric structure and electromagnetic field distribution on the axis of the designed 1.3 GHz 9-cell symmetrical twin-coupler cavity are shown in Fig.\ref{fig12}.
In this way, there are only two symmetrical fundamental-power couplers with twin-coupler cavity to ensure the symmetry of the structure, and there are no dipole field transverse components at both upstream and downstream coupler regions from Fig.\ref{fig12} (bottom plot). Then repeating the methodology of Section.\ref{sec.III} obtains the $E_z$ field at r = 3 mm downstream coupler position of the symmetrical twin-coupler cavity@$Q_e$=$4.12 \times 10^{7}$$\&$$Q_0$ = $2.7 \times 10^{10}$ and the normalized multipole field expansion coefficient $E_{m}/E_{0}$, and the results are shown in Fig.\ref{fig13} and Table.\ref{tab4}. From Fig.\ref{fig13} and Table.\ref{tab4}, it can be seen that the $E_z$ field at the twin-coupler has two cosine function periods, and the quadrupole mode become dominant mode. Compared with Fig.\ref{fig5} and Table.\ref{tab2}, the dipole mode coefficient of the symmetrical twin-coupler cavity is 5 orders of magnitude smaller than that of the TESLA cavity. Combining Fig.\ref{fig12}, Fig.\ref{fig13}, and Table.\ref{tab4}, we can deduce that the coupler RF kick of the symmetrical twin-coupler cavity can be ignored, and the resulting emittance suffered has little effect.

\begin{table}[!htbp]
\caption{Multipole field expansion factor $E_{m}/E_{0}$ for the symmetrical twin-coupler cavity@$Q_e$ = $4.12 \times 10^{7}$.}
\label{tab:Normalised_Fourier_coefficients_Teslacavity}
\centering
\begin{tabular} {cc}
    \hline
    Mode & Value   \\ 
    \hline
    $E_{0}/E_{0}$ ($a_{0}/a_{0}$),monopole & 1   \\ 
    $E_{1}/E_{0}$ ($a_{1}r_{0}/a_{0}$),dipole &  $1.4 \times 10^{-7}$\\
    $E_{2}/E_{0}$ ($a_{2}r_{0}^{2}/a_{0}$),quadrupole & \boldmath $5.3 \times 10^{-5}$\\
    $E_{3}/E_{0}$ ($a_{3}r_{0}^{3}/a_{0}$),sextupole &  $9.7 \times 10^{-8}$ \\
    $E_{4}/E_{0}$ ($a_{4}r_{0}^{4}/a_{0}$),octopole &  $1.2 \times 10^{-6}$\\
    \hline
    \label{tab4}
\end{tabular}
\end{table}

\begin{figure}[!htb]
    \centering
    \includegraphics*[width=0.7\linewidth]{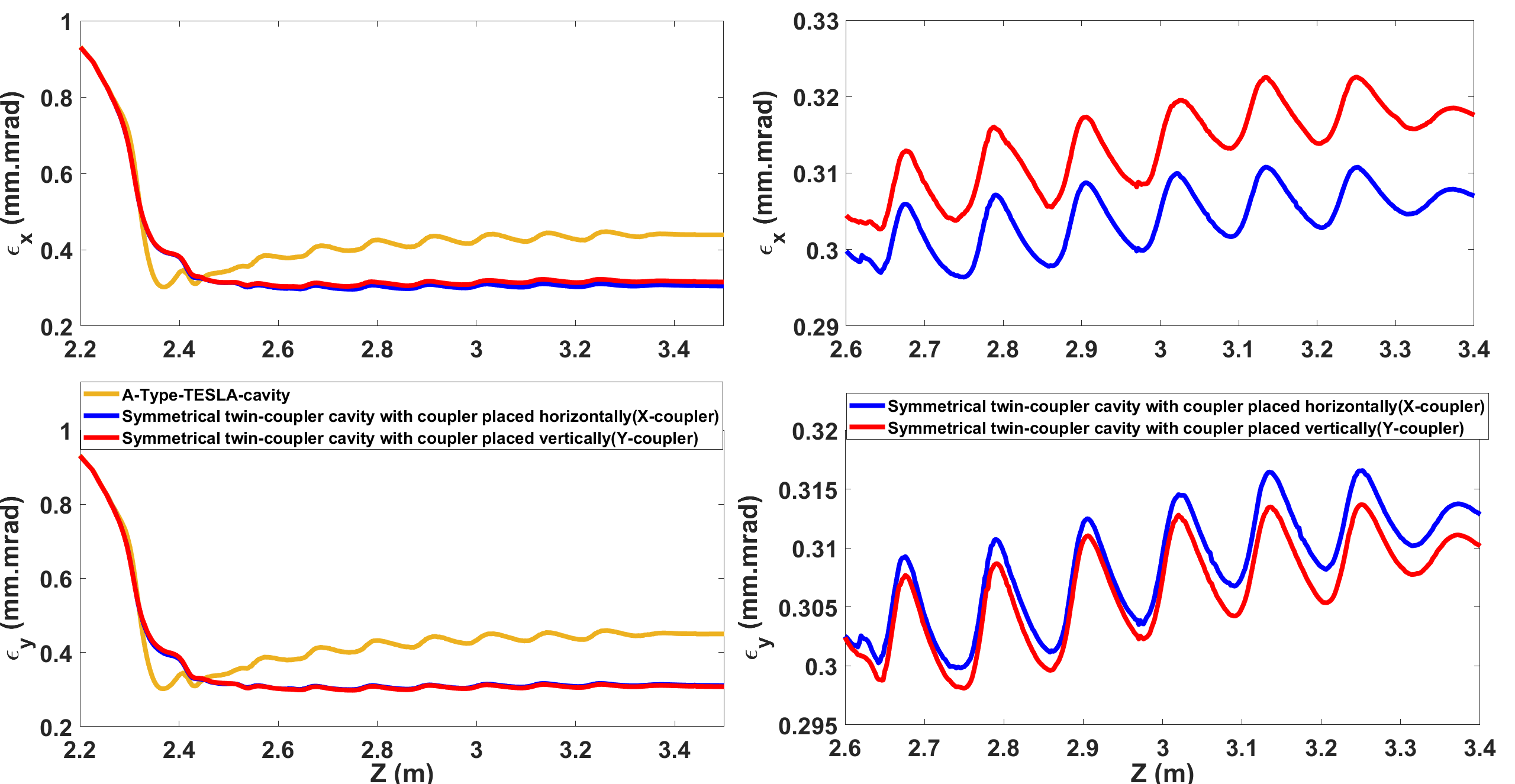}
    \caption{$\epsilon_x$ (top left) and $\epsilon_y$ (bottom left) of A-type, X and Y twin-coupler cavity, zoomed $\epsilon_x$ (top right) and $\epsilon_y$ (bottom right ) of X and Y twin-coupler cavity.}
    \label{fig14}
\end{figure}

In order to verify the above reasoning, we simulated the emittance of the injCM01 using Astra(Particle trajectory tracking program)~\cite{b28} at the working point in Table.\ref{tab1}, which imports the 3D field distribution for the following three cases:  a TESLA 9-cell cavity of the fundamental power coupler in the X-axis direction (called A-type TESLA cavity), 
the symmetrical twin-coupler cavity with coupler placed horizontally(X-coupler) and vertically(Y-coupler), and the result are shown in Fig.\ref{fig14}. 
As can be seen from such figure, it is clear that the emittance growth of the symmetrical twin-coupler cavity is much lower than that of A-type TESLA cavity, 
while $\epsilon_x$ of the symmetrical twin-coupler cavity with the coupler placed horizontally(X-coupler) is slightly less than that of vertically placed ones(Y-coupler), and the $\epsilon_y$ is about the same for both.

\begin{figure}[!htb]
   \centering
   \includegraphics*[width=0.7\linewidth]{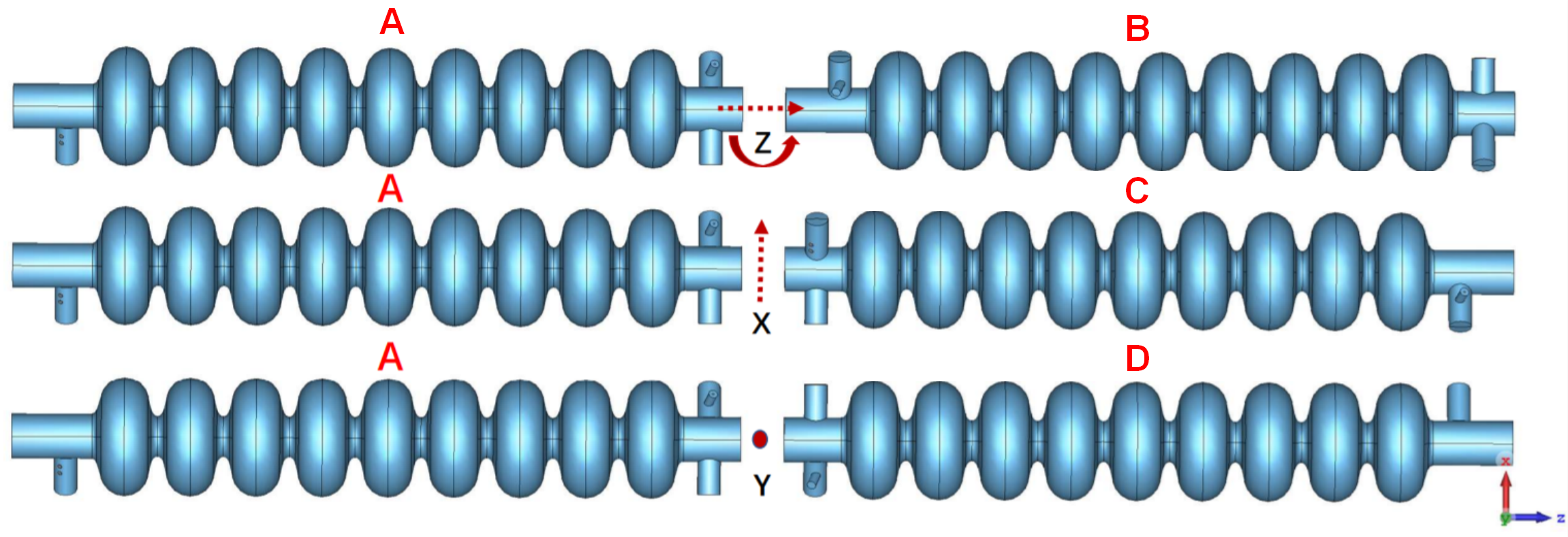}
   \caption{Four TESLA cavity types: ABCD.}
   \label{fig15}
\end{figure}

\begin{figure}[!htb]
   \centering
   \includegraphics*[width=0.7\linewidth]{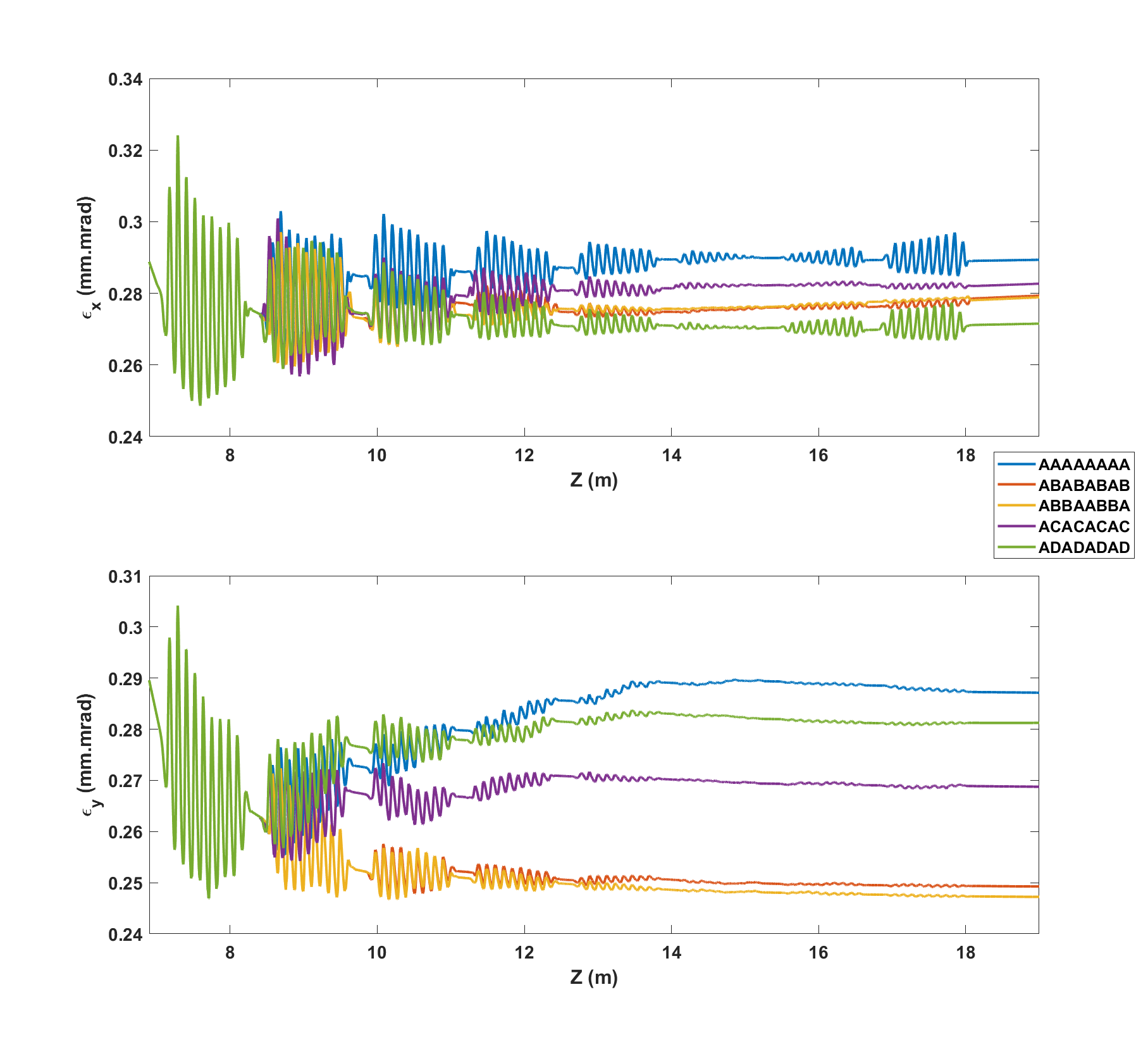}
   \caption{$\epsilon_x$ (top) and $\epsilon_y$ (bottom) for 5 permutation cases in the 8-cavities cryomodule(injCOM2).}
   \label{fig16}
\end{figure}

In the injCM02 of SHINE injector, due to the coupler of the 1.3 GHz 9-cell TESLA type cavity is directional, the rotation angle between neighboring cavities in the injCM02 can be reasonably arranged to reduce the emittance growth. As can be seen from Fig.\ref{fig10} and Table.\ref{tab3}, the ${\mathbf{K}}_{x}$ is much larger than ${\mathbf{K}}_{y}$ when the fundamental power coupler(FPC) of the TESLA cavty is placed horizontally(X direction), and electron bunch will deflect in the x direction when passing through the cavity, so emittance growth can be reduced by changing the rotation angle of the 8 TESLA cavities in the injCM02. 
We define 4 cavity types based on the rotation angle of the TESLA cavity, which are interpreted as follows:
\begin{itemize}
    \item [1)] A-Type TESLA cavity: TESLA cavity of the  fundamental power coupler in the x-axis direction
    \item [2)] B-Type TESLA cavity: A-Type TESLA cavity is rotated 180 degrees along the z-axis
    \item [3)] C-Type TESLA cavity: A-Type TESLA cavity is rotated 180 degrees along the x-axis
    \item [4)] D-Type TESLA cavity: A-Type TESLA cavity is rotated 180 degrees along the y-axis
\end{itemize}
the details as shown in Fig.\ref{fig15}.
Based on these 4 TESLA cavity types above, we have also defined 5 cases of different cavity permutations in the injCM02 as follows: (1) AAAAAAAA, (2) ABABABAB, (3) ABBAABBA, (4) ACACACAC, (5) ADADADAD. 
Finally, we compared emittance growth under these 5 cases using the ASTRA code, and the result are shown in Fig.\ref{fig16} and Table.\ref{tab5}.

From the Fig.\ref{fig16} and Table.\ref{tab5},  it is concluded as follows:
\begin{itemize}
\item  the emittance of AAAAAAAA case is the largest as expected
\item as the beam energy increases, the emittance growth gradually stabilizes after the fifth cavity (z = 14 m)
\item due to ${\mathbf{K}}_{x}$ $>$ ${\mathbf{K}}_{y}$ for the fundamental-power coupler in the x-direction, $\epsilon_x$ $>$ $\epsilon_y$
\item $\epsilon_y$ of the ABBAABBA case is much lower than the other 4 cases, while its $\epsilon_x$ is lower than that of the other 3 cases, and just higher than that of ADADADAD case
\item both $\epsilon_x$ and $\epsilon_y$ are smaller than 0.3 (much smaller than the design vaule of 0.4 in Table.\ref{tab1})
\end{itemize}

In summary, 1.3 GHz symmetric twin-coupler cavity with the coupler placed horizontally is adoped in the single-cavity cryomodule(injCM01) and ABBAABBA case is chosen in the 8-cavities cryomodule(injCOM2) respectively.  

\begin{table}[!htbp]
\caption{Emittance for 5 permutation cases at the end of the 8-cavities cryomodule(injCOM2).}
\label{tab5}
\centering
\begin{tabular} {ccc}
    \hline
    Cases & $\epsilon_x$(mm$\cdot$mrad) & $\epsilon_y$ (mm$\cdot$mrad)   \\ 
    \hline
    AAAAAAAA & 0.290 & 0.288  \\ 
    ABABABAB & 0.279 & 0.249\\
    ABBAABBA & \bfseries0.278 & \bfseries0.246\\
    ACACACAC & 0.282 & 0.268\\
    ADADADAD & 0.270 & 0.282\\
    \hline
    \label{tab5}
\end{tabular}
\end{table}

\section{Conclusion and discussion}\label{sec.VI}

In this paper, we discuss that the asymmetry of the superconducting cavity structure caused by the coupler, RF wave coupled into the TESLA cavity from the fundamental-power coupler, which generates a transverse dipole mode component (coupler RF kick) at the coupler region. A novel approach is provided to obtain the electromagnetic field distribution of the high-$Q_0$ superconducting cavity using lossy material simulations in the CST frequency domain solver, which is processed to derive the multipole field expansion coefficient($E_{m}/E_{0}$), and to calculate the coupler RF kick.
In order to minimize or even eliminate the dipole mode component and coupler RF kick of the injCM01 in the SHINE injector, a symmetric twin-coupler cavity has been designed so that dipole mode are no longer the dominant multipole mode, as well as the electromagnetic field on the axis in the coupler region no longer has a transverse component, which ultimately results in the symmetric twin-coupler cavity having a much lower emittance than the TESLA 9-cell cavity.
In the injCM02 of SHINE injector, the rotation angle and permutation of the 8 TESLA cavities can be optimized to decrease the beam emittance.
The ABBAABBA case of optimal emittance is obtained by comparing 5 cases of different cavity permutations in the injCM02.
Finally, a single-cavity cryomodule(injCM01) with a coupler horizontally placed twin-coupler cavity, and an 8-cavities cryomodule(injCM02) with an ABBAABBA case are adoped in the SHINE injector. 

\section*{Acknowledgements}

We are grateful to Dr. Yan Wang and Dr. Zhen-Yu Ma of Shanghai Advanced Research Institute, CAS, for his help with the twin-coupler cavity and the fundamental-power coupler of TESLA cavity, and many of the staff members and students at the SHINE and SXFEL project for their support with ideas and discussions.





\end{document}